\documentclass[a4paper,11pt]{article}
\usepackage{jcappub}
\usepackage[T1]{fontenc} 
\usepackage{color}

\usepackage{pdflscape}
\usepackage{amsmath}
\usepackage{amssymb}
\usepackage{dcolumn}
\usepackage{bm}
\usepackage{color}
\usepackage{epsfig}
\usepackage{amsfonts}
\usepackage{graphicx}
\usepackage{subfigure}
\usepackage{dcolumn}
\usepackage{graphicx}
\begin{document}

\title{Superposition of FLRW Universes}


\author[a]{Metin Gürses,}
\author[a]{Yaghoub Heydarzade,}
\author[b]{Bayram Tekin}


\affiliation[a]{Department of Mathematics, Faculty of Sciences, Bilkent University, 06800 Ankara, Turkey}
\affiliation[b]{Department of Physics,  Middle East Technical University, 06800 Ankara, Turkey}
\emailAdd{gurses@fen.bilkent.edu.tr}
\emailAdd{yheydarzade@bilkent.edu.tr}
\emailAdd{btekin@metu.edu.tr}

\abstract{We show that (1) the Einstein field equations with a perfect
fluid source admit a nonlinear superposition of two distinct homogenous
Friedman-Lemaitre-Robertson-Walker (FLRW) metrics as a solution, (2) the
superposed solution is an inhomogeneous geometry in general, (3) it
reduces to a homogeneous one in the two asymptotes which are the early and
the late stages of the universe as described by two different FLRW
metrics, (4) the solution possesses a scale factor inversion symmetry  and
(5) the solution implies two kinds of topology changes: one during the
time evolution of the superposed universe and the other occurring in the
asymptotic region of space.
}
\maketitle
\flushbottom
\section{Introduction}

An exact solution of the Einstein-perfect fluid system, representing an
inhomogeneous cosmological model and admitting various interpretations,
was recently rediscovered in \cite{our} by the method of separation of variables.
This solution is a subcase of the Kustaanheimo-Qvist class \cite{kust}, and coincides
with the other previously found solutions.
See \cite{kust} and \cite{krasin, bol} for the classification and interpretations of the previously found inhomogeneous
cosmological solutions. One of the possible interpretations  is that it is
a generalization of the McVittie solution representing a black hole immersed in a FLRW universe \cite{bol,feroni2,vaidya,blackhole}.    In this work we propose a different interpretation of our solution. The inhomogeneous cosmological model of \cite{our}
 is a nonlinear superposition of generically two distinct FLRW universes with different spatial curvatures. Nonlinear superposition in general relativity has been studied in the context of  B{\" a}cklund transformations \cite{chinea1,
chinea2, gur7,gur8} for  spacetimes that possess two Killing vector fields. Here we give a novel example of a nonlinear superposition in general relativity. The FLRW phases show up in the early and late stages of the universe. Since each FLRW metric has a different spatial curvature, then in the early and late eras of the universe the spatial curvatures are different in general
and a topology change is possible.  The superposed model has a scale factor
inversion symmetry akin to the scale factory duality symmetry observed  earlier in string cosmology \cite{dual}.

The layout of the paper is as follows. We first recapitulate the field equations of the inhomogeneous model in Section 2.
In Section 3, we introduce a solution to the field equations possessing the
scale factor inversion symmetry. In Section 4, we give two distinct interpretations
of the obtained solution. We mainly discuss the second interpretation: superposition
of two FLRW universes. In Section 5, we address two examples and discuss  the features of  our solution
in the context of  Big Bang  and  cyclic cosmological
models \cite{cyclic, ijjas, ccc}.  Section 6 is devoted to our conclusions.

\section{Matter-coupled field equations}
We consider the spherically symmetric metric in isotropic coordinates
\begin{equation}\label{*1c}
ds^2=-a^2(t,r) dt^2+c^4(t,r)\left(dr^2+r^2\,d\theta^2+r^2\, \sin^2\theta\, d\phi^2\right),
\end{equation}
where $a(t,r)$ and $c(t,r)$ are  generic differentiable functions of  time $t$ and  radial coordinate $r$. In \cite{our}, it was proven that the Einstein-perfect fluid field equations for $\dot c(t,r)\neq0$  reduce to
\begin{eqnarray}
&&2rc c^{\prime\prime}-6 r {c^{\prime}}^2-2c c^{\prime}- h(r)=0,\label{d}\\
&&a=2q(t)\,\frac{\dot c}{c},\label{a}
\end{eqnarray}
 where $h(r)$ is an arbitrary function of $r$ and
$q(t)$ is an arbitrary function of $t$; the fluid velocity is given as $u_\mu=a\delta^0_\mu$. Here a dot and a prime denote  differentiations with respect to $t$ and $r$, respectively. The energy density $\rho(t,r)$ and the pressure $p(t,r)$,
respectively read as
\begin{eqnarray}
&& 8 \pi\rho(t,r)=\frac{3}{q^2}-\frac{2}{ rc^{6}}\left(6r  {c^{\prime}}^2+6cc^{\prime}+ h \right)-\Lambda, \label{r}\\
&&8 \pi p(t,r)=-\frac{3}{q^2}+\frac{1}{ rq^3 c{^6} \dot
c}\, {\big(}2  q^3\,(-r {c^{\prime}}^2\dot c+rcc^{\prime}\dot c^{\prime} +cc^{\prime}\dot
c+c^2\dot c^{\prime})
+rc^7\dot q {\big)}
+\Lambda, \label{p}
\end{eqnarray}
where $\Lambda$ is the cosmological constant.  In \cite{our}, we discussed
that  if the metric functions $c(t,r)$ and $a(t,r)$ vanish on some surfaces then either $p$ or $\rho$  diverges, and hence  the Ricci  scalar    $R=8\pi(\rho-3 p)+4\Lambda$  diverges. These surfaces are defined as $\Sigma_{1}=\{(t,r)\in U | c(t,r)=0 \}$ and $\Sigma_{2}=\{(t,r) \in U | a(t,r)=0\}$ where $U$ is a part of spacetime and $0<t<\infty,~r \ge 0$. In particular, the surface $\Sigma_2$ represents namely the cosmological singularities \cite{penrose} or sudden cosmological singularities \cite{barrow1,barrow2, lake2004}.

\section{A solution to the field equations possessing the scale factor inversion symmetry }
In \cite{our},  it was shown that \begin{equation}\label{gj}
c(t,r)=\frac{\sqrt{R(t)}}{\sqrt{c_{0}+c_{1} r^2}}+\frac{\gamma}{ \sqrt{R(t)}} \frac{1}{\sqrt{c_2+c_{3} r^2}},
\end{equation}
is a solution to the  ordinary nonlinear differential
equation (\ref{d}) with arbitrary constants $c_{0}$, $c_{1}$, $c_{2}$, $c_{3}$ and $\gamma$; and   $R(t)$ is an at least  twice differentiable function.
   For this solution, $h(r)$ is given by
\begin{equation}
h(r)=\frac{6 \gamma (c_{0}\,c_{3}-c_{1}\,c_{2})^2 r^3}{(c_{0}+c_1 r^2)^{\frac{5}{2}}(c_2+c_3 r^2)^{\frac{5}{2}}},
\end{equation}
and furthermore choosing  $q(t)=R(t)/\dot R(t)$, the lapse function reads
as \begin{equation}
a(t,r)=\frac{1-\frac{\gamma}{ R(t)}\,\sqrt{\frac{c_{0}+c_{1}r^2}{c_{2}+c_3\, r^2}}}{1+\frac{\gamma}{R(t)}\, \sqrt{\frac{c_{0}+c_{1}r^2}{c_2+c_{3}r^2}}} \label{gfb}.
\end{equation}
Hence, the spacetime metric reads as
\begin{eqnarray}\label{*1c3}
&&ds^2=-\left(\frac{1-\frac{\gamma}{ R(t)}\,\sqrt{\frac{c_{0}+c_{1}r^2}{c_{2}+c_3\, r^2}}}{1+\frac{\gamma}{R(t)}\, \sqrt{\frac{c_{0}+c_{1}r^2}{c_2+c_{3}r^2}}} \right)^2 dt^2\nonumber\\
&&~~~~~~~~+\left(\frac{\sqrt{R(t)}}{\sqrt{c_{0}+c_{1} r^2}}+\frac{\gamma}{ \sqrt{R(t)}} \frac{1}{\sqrt{c_2+c_{3} r^2}} \right)^4\left(dr^2+r^2\,d\theta^2+r^2\, \sin^2\theta\, d\phi^2\right).
\end{eqnarray}
Substituting (\ref{gj}) in (\ref{r}) and (\ref{p}), one can find the asymptotic form when $R(t)\to 0$ and keeping the leading orders, $\rho(t,r)$ and $p(t,r)$ become homogeneous and reduce to\begin{eqnarray}
&&8\pi \rho(t)\to 3\frac{{\dot R}^2}{R^2}+\frac{12c_2 c_{3}R^2}{\gamma^4}-\Lambda,\label{1pis}\\
&&8\pi p(t)\to 2\frac{{\ddot R}}{R}-5\frac{{\dot
R}^2}{R^2}-\frac{4c_2c_3R^2}{\gamma^4}+\Lambda\label{1mis}.
\end{eqnarray}
Similarly, $\rho(t,r)$ and $p(t,r)$ also become asymptotically homogeneous as $R(t)\to \infty$  and read as\begin{eqnarray}
&&8\pi \rho(t)\to 3\frac{{\dot R}^2}{R^2}+\frac{12 c_0c_{1}}{R^2}-\Lambda,\label{11pis}\\
&&8\pi p(t)\to- 2\frac{{\ddot R}}{R}-\frac{{\dot
R}^2}{R^2}-\frac{4c_0c_1}{R^2}+\Lambda\label{11mis}.
\end{eqnarray}
Observe that the metric, i.e  (\ref{gj}) and (\ref{gfb}),  is invariant under the scale factor
 inversion, as $R\to \gamma^2/R$, $c_0\leftrightarrow c_2$ and $c_1\leftrightarrow c_3$. This symmetry is akin to the scale factor duality symmetry in string cosmology \cite{dual}.   This inversion symmetry  can also be observed  in the field equations:
under this symmetry, the pair of equations (\ref{1pis}) and (\ref{1mis})
go to (\ref{11pis}) and (\ref{11mis}) and vice versa.
\section{ Interpretations of  the solution}
\subsection{ A generalized McVittie metric}
Without losing any generality, one can choose the arbitrary constants $c_0, c_1, c_2, c_3$ and $\gamma$
in  such a way that the function $c(t,r)$ becomes
\begin{equation}\label{bszz}
c(t,r)=\frac{\sqrt{R(t)}}{\sqrt{\mu+ r^2}}+\frac{M}{ 2\sqrt{R(t)}} \frac{1}{\sqrt{1+k r^2}},
\end{equation}
with three constants $M, k$ and $\mu$ representing the  mass, spatial curvature of the background FLRW universe and the reduction parameter, respectively
and $R(t)$ is the same function as in (\ref{gj}) up to a multiplicative constant.
Letting $\mu=0$, the solution reduces to the uncharged Vaidya-Shah and the McVittie solution \cite{mcvittie,
vaidya}, see also \cite{krasin}. The McVittie
solution can be  interpreted as a black hole in a positively curved FLRW universe \cite{feroni2, blackhole}.
Next we shall give another
interpretation of this metric.
\subsection{Superposition of two FLRW universes}
 One can arrange the constants $c_{0}, c_{1}, c_{2}$ and $c_{3}$ in (\ref{gj}) so that  $c(t,r)$ takes the form
 \begin{equation}\label{gj1}
c(t,r)=\frac{\sqrt{R(t)}}{\sqrt{1+k_{1} r^2}}+\frac{\gamma}{ \sqrt{R(t)}} \frac{1}{\sqrt{1+k_{2} r^2}},
\end{equation}
where  $k_{1}$, $k_{2}$ and $\gamma$ are arbitrary constants and and $R(t)$ is the same function as in (\ref{gj}) up to a multiplicative constant.

To show that the two FLRW metrics are nonlinearly superposed, we need to use the FLRW metric in the isotropic coordinates which reads
\begin{equation}\label{**frw1}
ds^2=-dt^2+\frac{R^2(t)}{(1+k r^2)^2}\left(dr^2+r^2\, d\Omega^2\right),
\end{equation}
where $R(t)$ is the scale factor and $k$ is 1/4 of the curvature constant of 3-space (lets call it $k_*$); and after making the coordinate $r$ unitless, $k_*$ can  take the values $\pm 1$ and $0$. To discuss that our new solution represents the superposition of two FLRW universes, we consider the following two different FLRW
universes in the isotropic coordinates $(t, r, \theta, \phi)$.
\begin{description}
\item[${\bf FLRW_1}$:]\hfill \break
The metric for this case is given by
\begin{equation}
ds^2=-dt^2+c_1^4(t,r)\left(dr^2+r^2\, d\Omega^2\right),
\end{equation}
where
\begin{equation}\label{mnn}
c_1(t,r)=\frac{\sqrt{R(t)}}{\sqrt{1+k_1 r^2}}.
\end{equation}
This metric represents a FLRW universe with the scale factor $R(t)$ and the spatial curvature $k_{*}=4k_1$. The homogeneous matter density and pressure profiles for this case read as
\begin{eqnarray}
8\pi\rho_{1}(t)&=&3\frac{\dot R^2}{R^2}+
3\frac{ 4k_{1}}{R^2}-\Lambda,\label{00}\\
8\pi p_1(t)&=&-2\frac{\ddot R}{R}-\frac{\dot R^2}{R^2}-
\frac{ 4k_{1}}{R^2}+\Lambda.\label{01}
\end{eqnarray}
\item[${\bf FLRW_2}$:]\hfill \break
The metric for this case is given by
\begin{equation}\label{f2}
ds^2=-dt^2+c_2^4(t,r)\left(dr^2+r^2\, d\Omega^2\right),
\end{equation}
where
\begin{equation}\label{nmn}
c_2(t,r)=\frac{1}{\sqrt{R(t)}}\frac{1}{\sqrt{1+k_2 r^2}}.
\end{equation}
Then it represents a FLRW metric with the scale factor $ R^{-1}(t)$ and the spatial curvature $k_{*}=4k_2$.
Here, the matter density and pressure profiles are
\begin{eqnarray}
8\pi\rho_{2}(t)&=&3\frac{\dot R^2}{R^2}+
12k_{2} R^2-\Lambda,\label{11}\\
8\pi p_2(t)&=&2\frac{\ddot R}{R}-5\frac{\dot R^2}{R^2}-
4k_{2} R^2+\Lambda.\label{22}
\end{eqnarray}
\end{description}
Now, regarding the above two cases, we have the following theorem indicating
the second novel interpretation of the general solution (\ref{gj}) which we mainly discuss below.
\\

\noindent
\textbf{Theorem}: \textit{The linear superposition of $c_1(t,r)$ and $c_2(t,r)$, i.e. $c(t,r)=c_1(t,r)+\gamma c_2(t,r)$, with $c_1(t,r)$
and $c_2(t,r)$ given in (\ref{mnn}) and (\ref{nmn}) solves the Einstein field equations (\ref{d}), (\ref{a}), (\ref{r}) and (\ref{p}). Here, $\gamma$ is an arbitrary constant. Then,
this represents a nonlinear superposition of two particular solutions (\ref{gj1}) for which each part is generically a distinct} FLRW \textit{universe with  different spatial curvatures $k_1$ and $k_2$. If $k_1=k_2=k$, this general
 solution reduces to a single} FLRW \textit{solution with the spatial curvature $k_*=4k$.}
\\

The model has a Big Bang singularity if $R \to 0$ as $t \to 0$ without  further
assumption on the energy density and pressure, this is the only constraint on the function $R$. Hence, each choice of $R$ generates a different cosmological model. For instance, assuming the scale factor to be $R(t)=R_0\left(e^{\lambda t}-1 \right)^n$
where $R_0$, $\lambda$ and $n$ are positive constants, then from (\ref{1pis}) and (\ref{1mis}), one finds $\rho(t)\to \infty$ and $p(t)\to -\infty$ as $t \to 0$, respectively.
 In the limit $t\to \infty$, $R(t)\to
\infty$ and from (\ref{11pis}) and (\ref{11mis}), one finds $\rho(t)\to 3n^2
\lambda^2 -\Lambda$ and $p(t)\to -3n^2
\lambda^2 +\Lambda$, respectively,
 which corresponds to a (anti)de Sitter space. In this case, the universe starts from a Big Bang and evolves to a pure (anti)de Sitter phase at late times.

 Three consequences of the above theorem are as follows:  $(i)$ the
exact  solution unifies two different homogeneous FLRW solutions in a single superposed solution which generally is inhomogeneous,  $(ii)$ regarding
(\ref{1pis})-(\ref{11mis}), and  (\ref{00}), (\ref{01}), (\ref{11}) and (\ref{22}), the universe is approximately FLRW$_2$ as $R(t)\to 0$ and approximately FLRW$_1$
as $R(t)\to \infty$, see the item $(iv)$ in the next section for an expanding universe for more detail. Then the solution represents a phase transition from FLRW$_2$ to FLRW$_1$ with possibly a topology change:
If we denote the state of the universe with the FLRW parameters as $(R,k)$ where $R$ is scale factor and $k$ is the normalized curvature of 3-space.  According to our model, during the time evolution, the universe undergoes a change of state from $(1/R, k_{2})$ to a state  $(R, k_{1})$. In other words, the universe starts with a state $(1/R,k_{2})$ and ends with a different state $(R,k_{1})$. We call this change of state as a \textquotedblleft phase transition\textquotedblright,
and (iii) there exists a  scale factor inversion symmetry
in the metric  as $R \to \gamma^2/R$ and
 $k_{1} \leftrightarrow k_{2}$. Under this symmetry, FLRW$_2\leftrightarrow$ FLRW$_1$.

Let us expound on   the difference of the early and late eras (as $R \to 0$ and $R \to \infty$) of the universe as different FLRW universes. We can normalize only one of the parameters $k_{1}$ or $k_{2}$ by scaling the coordinate
$r$. By such a scaling $k_{1}$, $k_{2}$ and $r$ become unitless.
If the beginning of the universe is FLRW$_{2}$ with a normalized $k_{2}$ and ends as FLRW$_{1}$ with spatial curvature $k_{1}$, we have two possibilities, either sign ($k_{1}$) $=$ sign ($k_{2}$) (two universes having the same topology) and  sign ($k_{1}$) $ \ne$ sign ($k_{2}$) indicating a change of topology,
see the item $(iv)$ in the next section for an expanding universe for more detail.  In these asymptotic stages, the universe is approximately homogeneous.

In the following section, we elaborate on the properties of  the  solution for $k_1\neq k_2$ with two specific cosmological scenarios.

\section{Two specific cosmological scenarios}
\subsection{Expanding  universe scenario}
For an ever expanding universe with $R(t)\to 0$ as $t\to 0$ and $R(t)\to \infty$ as $t\to \infty$, or  a universe with  a minimum size $R_{min}~(R_{min}\ll1$) for $t\to 0$ and a maximum size $R_{max}~(R_{max}\gg1$) for $t\to \infty$,  we note the following  points.
\begin{description}
\item[$(i)$] In the early era  ($t\to 0$),  $c_2(t,r)$ dominates in $c(t,r)$. Then, the universe is effectively FLRW$_2$  and hence becomes homogeneous according to the matter density and pressure
given by  (\ref{11}) and (\ref{22}), respectively. See
the upper plot  in Figure \ref{fig1} for $k_1\geq0$ and $k_2>0$.
\item[$(ii)$] In the intermediate era ($0<t<\infty$), both $c_1(t,r)$ and $c_2(t,r)$ are effective  and hence we have a mixture of FLRW$_1$  and FLRW$_2$  universes which indeed is an inhomogeneous universe according to the matter density and pressure given by  (\ref{r}) and (\ref{p}), respectively (inhomogeneity can also be seen from any of the non-vanishing
curvature invariants such as the Ricci scalar which has position
dependence). In this case, cosmological inhomogeneities emerge and contribute
to the formation
of structures in the universe. See
 Figure \ref{fig1} for $k_1\geq0$ and $k_2>0$
 where the plots A and C represent initial
and final homogeneous FLRW universe while the plot B indicates a mixed inhomogeneous
universe. The dots in the plot B depict symbolically the inhomogeneities in the mixed phase.
\item[$(iii)$]  In the late times ($t\to\infty$),   $c_1(t,r)$ dominates in $c(t,r)$. Therefore, the universe tends effectively
to FLRW$_1$  and hence becomes homogeneous according to (\ref{00}) and (\ref{01}). See the upper plot in Figure \ref{fig1} for $k_1\geq0$ and $k_2>0$.
\begin{figure}[h!]
\begin{center}
\includegraphics[scale=1]{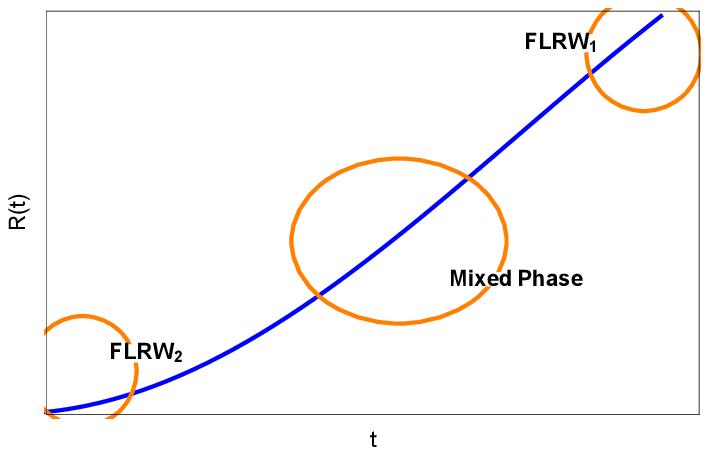}
\includegraphics[scale=1]{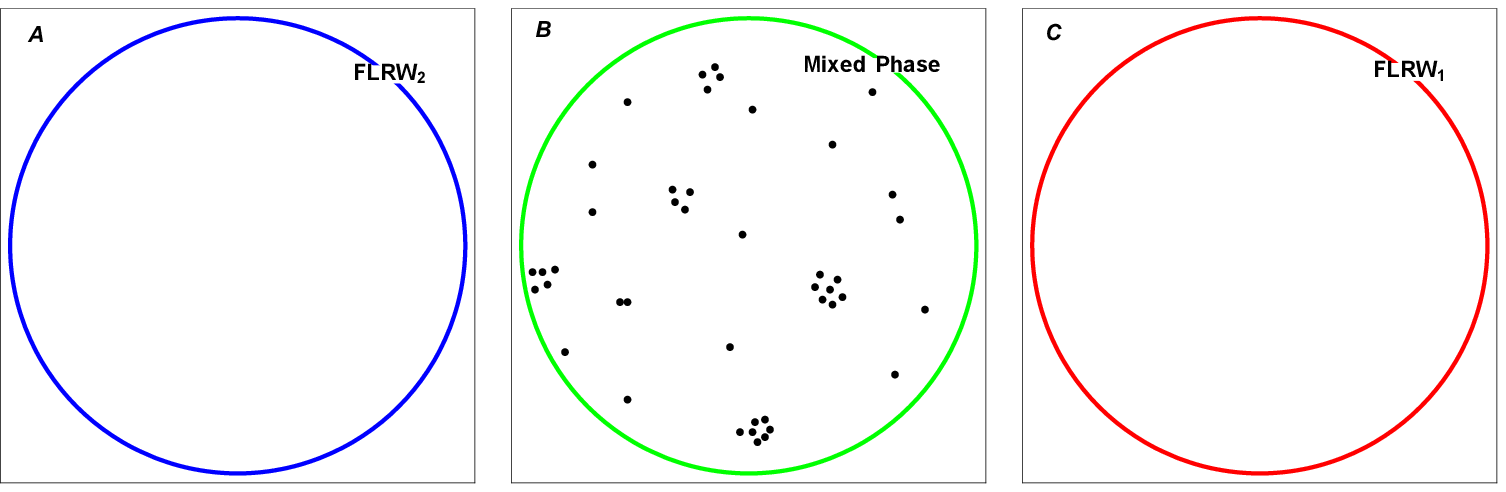}
\caption{The upper plot represents two different initial and final FLRW universes.
The plots  A and C represent two different initial and final homogeneous
universes while the plot  B denotes growing inhomogeneities in the intermediate mixed phase.  \label{fig1} }
\end{center}
\end{figure}
\item[$(iv)$] An important consequence of the phase transition between FLRW$_2$  and FLRW$_1$ is the topology change in the universe. Since  FLRW$_2$  dominating at early times and FLRW$_1$ dominating at late times have different spatial curvatures in general,  the universe may undergo
a topology change from $k_2$ to $k_1$ during its evolution. For this  topological
transition, one realizes the following two points.
\begin{description}
\item[$1.$]
If $k_1\geq0$ and $k_2>0$ (or $k_1>0$ and $k_2\geq0)$, hence  $r\in[0, \infty)$.
This case represents the topology change from a closed universe to another flat
or closed universe (or vice versa).
To observe this topology change, considering $k_1\geq0$ and $k_2>0$ in
(\ref{gj1}), we have
\begin{equation}\label{gjgt}
c(t,r)=\frac{\sqrt{R(t)}}{\sqrt{1+k_1r^2}}+\frac{\gamma}{ \sqrt{R(t)}} \frac{1}{\sqrt{1+k_{2} r^2}}.
\end{equation}
The dominant term in $c(t,r)$ at the asymptotic limit $t\to 0$ and thus $R(t)\to 0$ is the second term with the topology $k_2>0$. As time progresses toward the late times, i.e as $t\to \infty$ and thus $R(t)\to \infty$, the first term in (\ref{gjgt}) with the topology  $k_1\geq0$ dominates and consequently a change of topology occurs.
\item[$2.$] If $k_1\geq0$ and  $k_2<0$ or ($k_1<0$ and $k_2\geq0$). This case represents a topology
change from a spatially open to a flat or closed universe (or vice versa).
 To observe the topology
change in this case, we have
\begin{equation}\label{gjkls}
c(t,r)=\frac{\sqrt{R(t)}}{\sqrt{1+k_{1} r^2}}+\frac{\gamma}{ \sqrt{R(t)}} \frac{1}{\sqrt{1-|k_{2}| r^2}}.
\end{equation}
Here, there is a restriction on the coordinate patch   as $r\in [0, \frac{1}{\sqrt{|k_2|}})$.
If $k_1=0$, in the asymptotic limit $t\to 0$ and hence $R(t)\to 0$, regardless
of whatever $r$ is, the second term in (\ref{gjkls}) dominates. Then, at the early time asymptotic state, the topology is $k_2<0$.
At the late times, there are two possibilities as $(i)$ in the region $0\leq
r\ll\frac{1}{\sqrt{|k_2|}}$, as $t\to \infty$ and $R(t)\to \infty$, the first term in (2) dominates and a topology change  occurs in time. In this region,
the universe starts with the open topology $k_2<0$ and evolves toward the
state with a flat topology
$k_1=0$, and $(ii)$ in the asymptotic region, i.e. $r\to \frac{1}{\sqrt{|k_2|}}$, the second term  survives in the limit $t\to \infty$ and $R(t)\to \infty$. Thus,  both the terms in (\ref{gjkls}) can be effective that preserves the inhomogeneity at the asymptotic
region. For this case, the topological structure of the asymptotic region can be more complicated than the previous cases and it may  be different than both
of the $k_1$ and $k_2$. It is interesting that here  the topologies for the internal and asymptotic regions of spacetime  can be different which  implies another kind of topology change. A similar topology change occur also for $k_1<0$ and $k_2\geq0$.
In particular, the case $k_1<0$ and $k_2>0$ represents the topology change of the spacetime  akin to the "bag of gold" geometry  of Wheeler \cite{wheeler}.
\end{description}
\end{description}
\subsubsection*{A Particular Example: Initially inflating and finally accelerating expanding universe scenario}
\begin{figure}[ht]
\begin{center}
\includegraphics[scale=0.78]{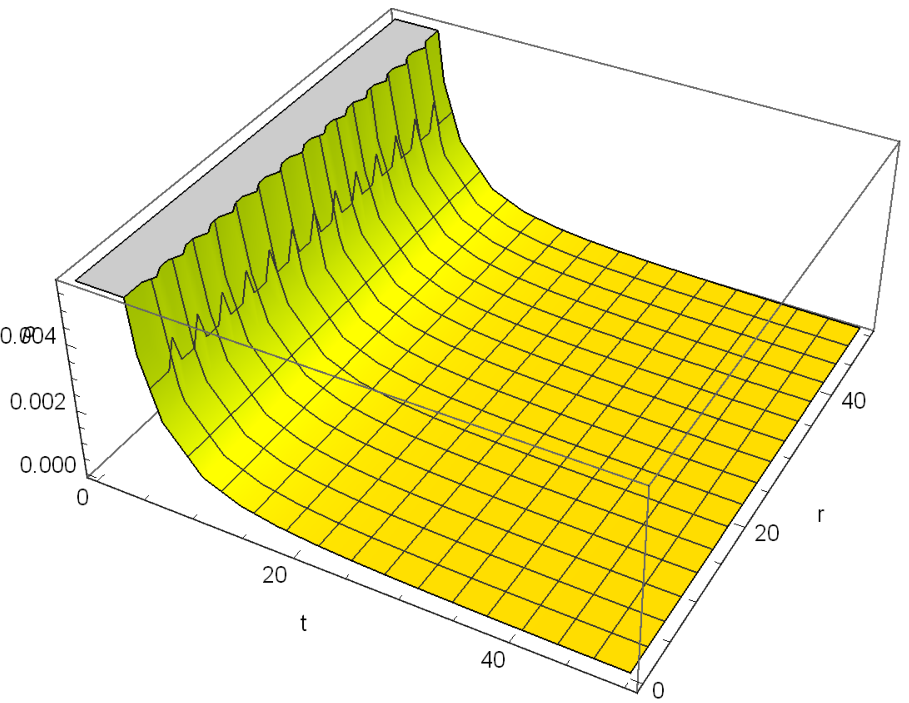}
\includegraphics[scale=0.78]{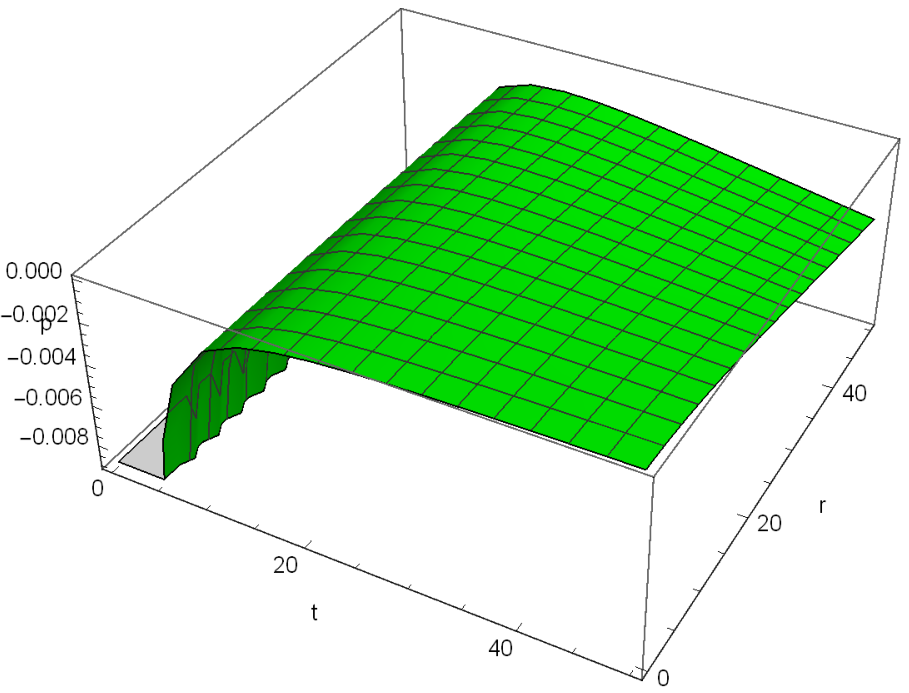}
\includegraphics[scale=0.78]{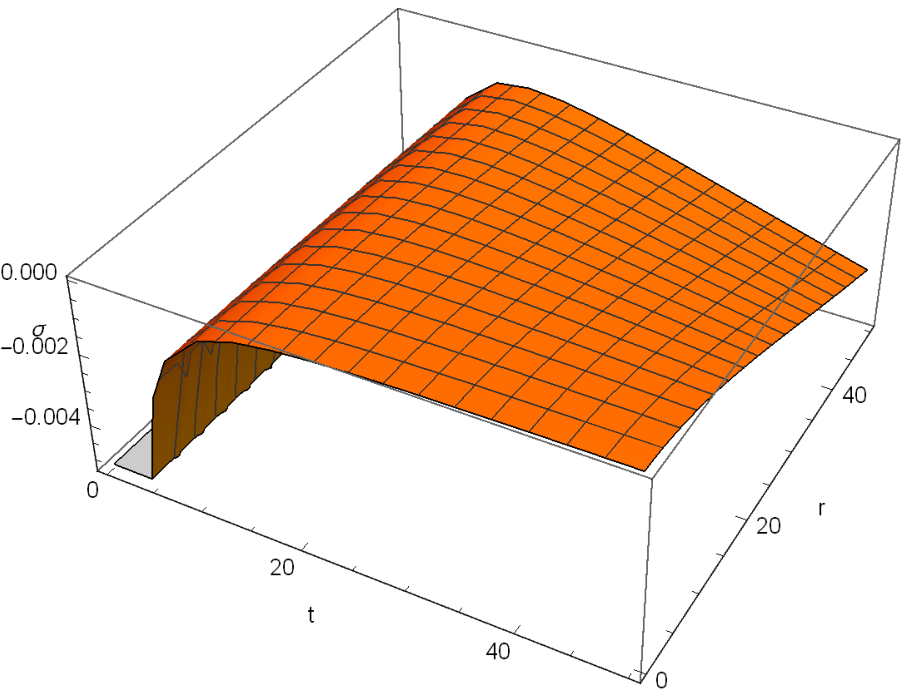}
\caption{The evolution of the density $\rho(t,r)$, pressure $p(t,r)$ and $\sigma(t,r)=\rho+p$, respectively, for the parameter values $k_1 = 0,\, k_2 = 1,\, \gamma=1,\, R_0=1,\, \sqrt{\Lambda/3}= 0.0001$,
and $R(t) = R_0 \left( e^{ \sqrt{\frac{\Lambda}{3}} t} - 1\right)$.
\label{den}}
\end{center}
\end{figure}

Considering the scale factor
$R(t)=R_0 \left(e^{\sqrt{\frac{\Lambda}{3}}t}-1 \right)$,
one observes that in the early universe as $R(t)\to 0$, FLRW$_2$ (\ref{f2}) dominates in the superposition. Then, the evolution of  the universe is governed by (\ref{11}) and (\ref{22}). According to (\ref{11}) and (\ref{22}), we
have $\rho_2(t)\to \infty$ and $p_2(t)\to -\infty$ as $t\to 0$, respectively. The first  represents the initial  singularity in
the matter sector while the latter represents a self driven inflation in the early times.  It the late times,  as $t\to \infty$, $R(t)\to R_0e^{\sqrt{\frac{\Lambda}{3}}t}$,  FLRW$_1$ (\ref{f2}) is dominant in the superposition and the  universe evolves according to (\ref{00}) and (\ref{01})
to an accelerating expanding a pure de Sitter phase, i.e. $\rho_1(t)\to 0$ and $p_1(t)\to 0$ as $t\to \infty$. Then, one observes
an interesting property
in this exact solution to the Einstein field equations. Indeed, this
solution provides a scenario including the early time inflation, inhomogeneous
structure
formation in the intermediate era, and the late time accelerating expansion in a unified model possessing a possible topology change.
In the intermediate phase where both FLRW$_1$ and FLRW$_2$ contribute effectively
to the evolution of the universe, the inhomogeneities emerge. This possibility
cannot be achieved in the usual  standard FLRW cosmology
with only  time dependent field equations. One may argue about the physical
nature of the solution by addressing the matter density, pressure and energy conditions. In Figure \ref{den}, we have plotted  the  density $\rho(t,r)$
and pressure $p(t,r)$ in (\ref{r}) and (\ref{p}), respectively, as well as $\sigma(t,r)=\rho+p$ for some typical values of the parameters. It is seen that for a superposed universe undergoing a topology change from a closed to flat topology with the scale factor $R(t)=R_0\left(e^{\sqrt{\frac{\Lambda}{3}}t}-1 \right)$, there are regions where the density remains
positive and pressure and $\sigma$ are negative.  The negativeness of  $\sigma$ represents the violation of weak energy condition that is  consistent with the expanding nature of the cosmos in the context of this solution.
\subsection{Cyclic universe scenarios}
\begin{figure}[ht]
\begin{center}
\includegraphics[scale=1]{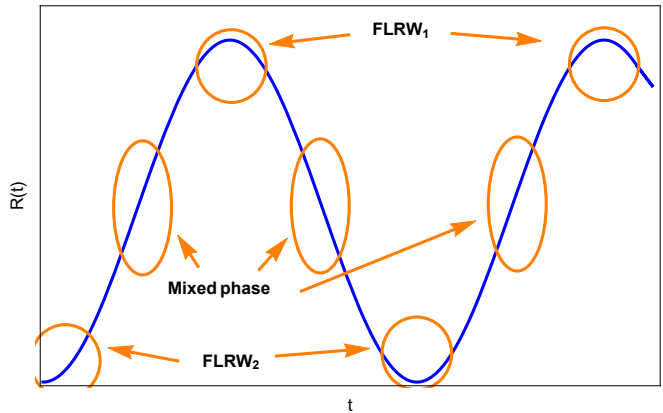}
\caption{The evolution of a cyclic universe undergoing
a topology change during the oscillation between  subsequent global  maximum and minimum values
of its scale factor $R(t)$.\label{oscillation1}}
\end{center}
\end{figure}

There are various cyclic universe scenarios,
as exemplified in \cite{cyclic, ijjas, ccc}. The old cyclic scenario  is based on the possibility that the scale factor $R(t)$ of the
universe oscillates at regularly spaced intervals of time between maximum
and minimum values \cite{cyclic}.  Considering such a scale factor,   the superposed
universe may undergo a topology change during each of its expansion and contraction phases between those minimum and maximum values of the scale factor $R(t)$,  see Figure \ref{oscillation1}. As the scale factor reaches its minimum and maximum values, the universe becomes approximately homogeneous FLRW$_2$ and FLRW$_1$, respectively, and it is in the mixed phase between these extremum
points  where
inhomogeneities emerge.

\begin{figure}[ht]
\begin{center}
\includegraphics[scale=0.5]{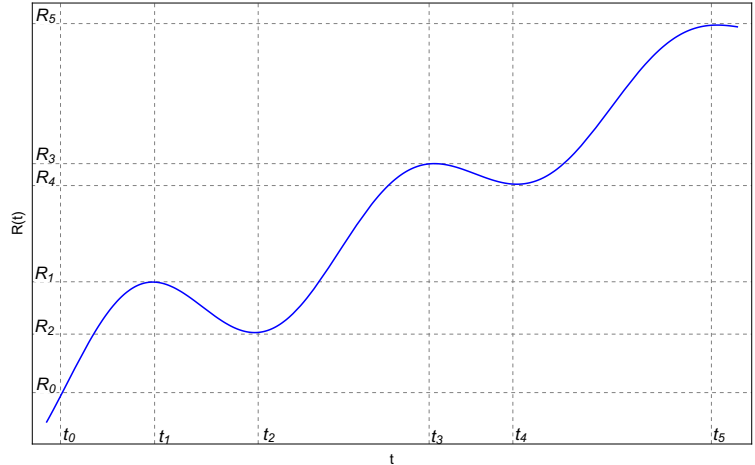}\\
\caption{The evolution of a cyclic universe undergoing
a topology change in transition through the local maximum and minimum values
of a typical scale factor $R(t)=e^{\lambda t}J^2_0(t) + 4$ where  $J_0$ is the zeroth order Bessel function and we have set the dimensionless parameter
$\lambda=0.009$. \label{oscillation2}}
\end{center}
\end{figure}

Another possible type of cyclic universe scenario was proposed recently by Ijjas and Steinhardt \cite{ijjas}. In this model, instead of the scale factor $R(t)$, the Hubble parameter
$H(t)=\dot R(t)/R(t)$  oscillates periodically during the evolution of
the universe. In the context of this model, the scale factor
$R(t)$ increases substantially during each cosmological era and then undergoes an  ultra-slow contraction phase at the end of each cycle. Then,
the next
 cycle of the universe begins with a non-singular bounce.  Figure \ref{oscillation2}
represents a typical plot of this type of cyclic universe scenario. In the first cycle, from $t_0$ to $t_2$, the universe expands from FLRW$_2$
at $t_0$ with the spatial curvature $k_2$ to FLRW$_1$ at the local maximum of $R(t)$ at $t_1$ with the spatial curvature of $k_1$. Then, it contracts between $t_1$ to $t_2$
and recover its previous curvature  $k_2$ of the FLRW$_2$ state provided that $R_2\ll R_{1}$. In both the expansion and contraction phases, the universe becomes
inhomogeneous in between the local maximum and minimum values of the scale factor. One also notes that by flattening the contraction phases, i.e. in the second
cycle where $R_3\sim R_4$, the universe keeps its topology in the contraction
phase as in the past local maximum point but still evolves inhomogeneously
toward the non-singular bouncing point.

As a final remark, let us  note that although in Penrose's conformal cyclic cosmological picture \cite{ccc} there is no contraction phase, since in each aeon the universe is ever expanding ($R(t_{f})\gg R(t_i)$ where $i$ and $f$ denote the initial and final states of an arbitrary
aeon)  the universe undergoes a  change of spatial curvature from $k_2$ to $k_1$ during the evolution of each aeon. This requires a sudden  change of
spatial curvature and possibly topology change at the transition point from  the past aeon to the present aeon  from $k_1$ to $k_2$. This sudden transition point corresponds to the Big Bang of the present aeon. However,
since the metric of the past aeon at its null infinity and the metric of the present aeon
at its Big Bang surface are conformally related, occurrence of such a sudden topology
change is forbidden in the conformal cyclic cosmology picture.

\section{Conclusions}
A  summary of what we propose in this work is as follows: (1) we give a new example of nonlinear superposition in general relativity. This is a superposition of two different homogenous FLRW universes yielding  an  inhomogeneous cosmological model. (2) The metric is invariant under  the
scale factor inversion. (3) If the scale factor $R$ is zero in the beginning of the universe and goes to infinity as $t \to \infty$ then the universe starts approximately as a FLRW universe and ends as a different FLRW universe. (4) During such a phase transition the
spatial curvature  of 3-space changes in both magnitude and sign. If the sign
changes then the topology of the 3-space also changes but if the sign remains intact, then the spatial curvature of the 3-space either increases or decreases.

\end{document}